\documentclass[aoas]{imsart}

\RequirePackage{amsthm,amsmath,amsfonts,amssymb}
\RequirePackage[authoryear]{natbib}
\RequirePackage[colorlinks,citecolor=blue,urlcolor=blue]{hyperref}
\RequirePackage{graphicx}

\startlocaldefs
\endlocaldefs

\begin{document}

\begin{frontmatter}

\title{Estimating Functional Parameters for Understanding the Impact of Weather and Government Interventions on COVID‐19 Outbreak}
\runtitle{Impact of Weather and Government Interventions on COVID‐19 Outbreak}

\begin{aug}
\author{\fnms{Chih-Li} \snm{Sung}\ead[label=e1]{sungchih@msu.edu}}
\address{Department of Statistics and Probability,
Michigan State University, \printead{e1}}
\end{aug}

\begin{abstract}
As the coronavirus disease 2019 (COVID-19) has shown profound effects on public health and the economy worldwide, it becomes crucial to assess the impact on the virus transmission and develop effective strategies to address the challenge. A new statistical model derived from the SIR epidemic model with functional parameters is proposed to understand the impact of weather and government interventions on the virus spread in the presence of asymptomatic infections among  eight metropolitan areas in the United States. The model uses Bayesian inference with Gaussian process priors to  study the functional parameters nonparametrically, and sensitivity analysis is adopted to investigate the main and interaction effects of these factors. This analysis reveals several important results including the potential interaction effects between weather and government interventions, which shed new light on the effective strategies for policymakers to mitigate the COVID-19 outbreak. 
\end{abstract}


\end{frontmatter}

\section{Introduction}
As the coronavirus disease 2019 (COVID-19) has already had profound effects on public health and the economy worldwide, how to use statistical approaches to model and understand the spread of COVID-19 to inform and educate the public about the virus transmission and develop effective strategies for addressing this challenge has become crucial. In particular, the understanding of how government interventions and environmental factors, such as temperature and humidity, affect the virus transmissibility is important yet unclear. Moreover, an effective strategy to mitigate the outbreak based on the weather conditions is in extreme need for policymakers yet little attention has been paid to the interaction effect between weather and government interventions. For instance, a natural question for policymakers is ``Should the government implement more restrictions to mitigate the pandemic as the weather gets colder?''

Since the COVID-19 outbreak, many studies have investigated the impact of weather and government interventions, but  some challenges remain. \cite{yu2020impact,xu2020modest,carson2020covid} find some evidence that the weather may be associated with the COVID-19 spread, while \cite{jamil2020no,gupta2020estimating} find no significant associations. Most of the studies on the impacts of government interventions on COVID-19 spread show that government interventions are associated with reduced COVID-19 transmission; e.g., \cite{cowling2020impact,haug2020ranking,haldar2020effect,flaxman2020estimating}. However, most of this work focuses on individual effects of weather and government interventions, which may lead to misleading results due to  potential collinearity issues \citep{wilson2020weather}. For example, cold weather may increase the risk of disease overall, leading governments to impose travel restrictions. Further, interaction effects between weather and government interventions variables cannot be estimated if the effects in these two sets of variables are estimated separately. Moreover, most of the existing work study the impact \textit{without} accounting for the presence of the persons who are asymptomatic but can nevertheless infect others, which is a distinguishing feature of COVID-19 in comparison with previous viral diseases.

In this paper, we employ a nonparametric regression method not only  to model the impacts of weather and government interventions \textit{jointly} in the presence of asymptomatic infections, which incorporates an epidemic model that allows us to evaluate the effects on the virus transmissibility, but also provide forecasts of future COVID-19 infections. Specifically, a Gaussian process prior \citep{williams2006gaussian} is imposed on the functional  parameters in the \textit{susceptible-infectious-removed (SIR)} model \citep{kermack1927contribution}, and based on this model, the posterior distribution of the basic reproduction number, which is used to measure the transmission potential of a disease, will be derived. Both main and interaction effects of these factors will be analyzed by sensitivity analysis \citep{sobol1993sensitivity}. 

Parameter estimation in epidemic models is often called \textit{calibration} in the computer experiment literature \citep{kennedy2001bayesian,santner2003design,tuo2015efficient}. Although there are numerous developments on calibration, most of the existing work are based on scalar parameters rather than functional parameters. Exceptions include the recent work by  \cite{plumlee2016calibrating,brown2018nonparametric}, but their work is based on continuous outputs with a Gaussian assumption, which does not hold for count data in the epidemic models.

In Section \ref{sec:sir}, the SIR model and a modified SIR model with functional parameters will be introduced. The statistical model incorporated with the SIR model will be explicitly described in Section \ref{sec:statistcalmodel}. Numerical studies are conducted in Section \ref{sec:simulation} to examine the performance. In Section \ref{sec:application}, the statistical model is applied to the COVID-19 outbreak to assess the impacts of weather and government interventions. Final remarks are given in Section \ref{sec:conclusion}. The details of sampling for the posterior distributions, the R \citep{R2018} code, and the data for reproducing the results in this paper are provided in the Supplementary Material \citep{suppsung2022}.

\section{Compartmental models in epidemiology}\label{sec:sir}
\subsection{SIR model}
Compartmental models are widely used in epidemiology which simplify the mathematical modelling of infectious diseases. One of the prominent models is the \textit{SIR model} \citep{kermack1927contribution,Diekmann2013}, which assigns the population to three compartments: susceptible (S), infectious (I), and removed (R), where the three compartments respectively represent the number of the susceptible individuals, the infected individuals, and the removed individuals, which include the ones who are recovered, quarantined or deceased. The SIR model has been widely used for understanding how a disease spreads in outbreaks of measles, influenza, rubella, smallpox, Ebola, monkeypox, SARS, and the current COVID-19 pandemic. See, for example, \cite{osthus2017forecasting,chen2020time,cooper2020sir,roda2020difficult,d2020assessment}. 

Transitions among the three compartments can be expressed mathematically by three ordinary differential equations as follows:
\begin{equation}\label{eq:sir}
    \frac{dS(t)}{dt}=-\frac{\beta I(t)S(t)}{N},\quad\frac{dI(t)}{dt}=\frac{\beta I(t)S(t)}{N}-\gamma I(t),\quad\frac{dR(t)}{dt}=\gamma I(t),
\end{equation}
where $S(t)$, $I(t)$ and $R(t)$ represent the numbers of cases in the corresponding compartments, $N=S(t)+I(t)+R(t)$ is the total population, $\beta$ is the contact rate that represents the average number of contacts per person per time in the susceptible compartment that is sufficient to spread the disease, and $\gamma$ is the removed rate from the infectious compartment to the removed compartment. The ratio of $\beta$ and $\gamma$ is called the \textit{basic reproduction number} in epidemiology, often denoted by $\mathcal{R}_0:=\beta/\gamma$, which indicates the average number of infected cases generated by a typical infectious individual when introduced into a fully susceptible population. This number is of great importance in public health and epidemiology, which is often used to measure the transmission potential of a disease or a virus \citep{dietz1993estimation,zhao2020preliminary,zhang2020evolving}. Essentially, when $\mathcal{R}_0$ is larger than 1, the infection will be able to start spreading in a population, and the larger $\mathcal{R}_0$ is,  the harder it is to control the epidemic.

\subsection{Modified SIR model}
Although the SIR model has been widely used in epidemiology, the model has been shown that it cannot reflect the reality due to its simplifications and assumptions. See, for example, \cite{ModelingInfectious,sung2020efficient}. In particular, the constant parameter assumption of $\beta$ and $\gamma$, which implies that the contact rate and the removed rate  are both fixed in the entire process, is too strong and unrealistic \citep{cowling2008effectiveness,cauchemez2016unraveling,hong2020estimation,yu2020impact,ambrosio2020coupled}. Therefore, in this article, we consider a modified, more flexible SIR model by assuming that the parameters can  vary based on potential factors. 

First, similar to \cite{hong2020estimation}, we  consider a discrete version of SIR models by replacing the derivatives in \eqref{eq:sir} with finite differences, which results in 
\begin{align*}
    I(t+1)&-I(t)=\frac{\beta I(t)(N-I(t)-R(t))}{N}-\gamma I(t),\\
     &R(t+1)-R(t)=\gamma I(t).
\end{align*}
Then, by assuming the functional parameters $\beta(\mathbf{x})$ and $\gamma(\mathbf{x})$, where $\mathbf{x}\in\Omega\subseteq\mathbb{R}^d$ is a $d$-dimensional factor, and expressing the equations in a recursive fashion, a modified SIR model can be expressed as 
\begin{align*}
    I(t+1)&=(1+\beta(\mathbf{x}) -\gamma(\mathbf{x}))I(t) + \beta(\mathbf{x})I(t)(I(t)+R(t))/N,\\
     &R(t+1)=R(t)+\gamma(\mathbf{x}) I(t),
\end{align*}
and $S(t+1)=N-I(t+1)-R(t+1)$ for $t\in\mathbb{N}\cup\{0\}$. Thus, the number of the daily infectious cases at day $t$ based on the modified SIR is the difference in susceptible from day $t-1$ to day $t$, which we denoted as
\begin{equation}\label{eq:modifiedSIR}
    f(t,\beta(\mathbf{x}),\gamma(\mathbf{x})):=S(t-1)-S(t).
\end{equation}

\section{A statistical model incorporated with the SIR model}\label{sec:statistcalmodel}
\subsection{Gaussian process priors for functional parameters}
In this section, we introduce a statistical model incorporated with the modified SIR model in \eqref{eq:modifiedSIR}. 
First, denote $y_t$ as the daily reported number of infectious cases at day $t$, and assume $y_t$ follows an independent Poisson distribution with the mean function that is a fraction of $f(t,\beta(\mathbf{x}), \gamma(\mathbf{x}))$. That is,
\begin{equation}\label{eq:dataassumption}
y_t\overset{\text{indep.}}{\sim}\text{Poi}(\kappa(t)f(t,\beta(\mathbf{x}),\gamma(\mathbf{x}))),
\end{equation}
where $\kappa(t)\in(0,1)$, indicating that only a fraction $\kappa(t)$ of the total number of infected are reported. The fraction $\kappa(t)$ plays a crucial role for taking into account the presence of asymptomatic, or however undetected, infectious cases, which is one of the distinguishing features of COVID-19. The idea of including the fraction was also mentioned in the literature (e.g., \cite{piazzola2021note} and \cite{ansumali2020modelling}), but a constant fraction was often considered, which was criticized by \cite{ansumali2020modelling} as unrealistic. A dynamic fraction as in \eqref{eq:dataassumption} varying with time is more realistic for the problem.  It should be noted that in the model \eqref{eq:dataassumption}, it is intrinsically assumed that the rate of transmission by an infected-asymptomatic person is the same as the infected-symptomatic person. More sophisticated models with an additional ``asymptomatic-but-infected'' compartment, which allow for different transmission rates in the model, may be considered, such as \cite{robinson2013model} and \cite{ansumali2020modelling}. However, as pointed out by \cite{ansumali2020modelling}, several recent studies show that there is no discernible difference between the two rates in the COVID-19 outbreak. See, for example, \cite{he2020temporal,li2020substantial,liu2020viral,wolfel2020virological}. As such, equal transmission rates are assumed in the proposed model, and the extension to the compartmental models with the asymptomatic-but-infected compartment, such as the SAIR model of \cite{ansumali2020modelling}, is left for the future work. Such work could also allow for studying whether the effects of potential factors differ among the asymptomatic versus symptomatic persons, such as government interventions.

Notably, as pointed out by  \cite{ansumali2020modelling}, another popular epidemic model for the outbreak, SEIR model (Susceptible-Exposed-Infectious-Removed) \citep{kermack1927contribution}, is not as realistic, because the exposed group (E) of the SEIR model does not infect the susceptible group (S) as the E group does not carry a sufficient viral load to infect others through contact. This is unlike the asymptomatic-but-infected individuals in the COVID-19 outbreak, which \textit{do} lead to the S group getting infected. As such, a model like the proposed model accounting for these asymptomatic individuals is more realistic.

The functional parameters in the SIR model are assumed to follow a joint Gaussian process (GP) prior:
\begin{align}\label{eq:GPassumption}
{\rm logit }\,\left(
\begin{array}{c}
\beta(\cdot)\\
\gamma(\cdot)\\
\end{array}\right)\sim\mathcal{GP}\left(\left[
\begin{array}{c}
\mu_1(\cdot)\\
\mu_2(\cdot)\\
\end{array}\right], \tau\mathbf{A}\left[
\begin{array}{cc}
K_{\boldsymbol{\phi}_1}(\cdot,\cdot)&0\\
0&K_{\boldsymbol{\phi}_2}(\cdot,\cdot)\\
\end{array}\right]\mathbf{A}\right),
\end{align}
where $\text{logit}(x)=\log \frac{x}{1-x}$ and
\begin{align}\label{eq:GPassumption_correlation}
    \mathbf{A}&=\left[
\begin{array}{cc}
1&\rho\\
\rho&1\\
\end{array}\right]^{1/2}=\frac{1}{\sqrt{2+2\sqrt{1-\rho^2}}}\left[
\begin{array}{cc}
1 + \sqrt{1-\rho^2}&\rho\\
\rho&1 + \sqrt{1-\rho^2}\\
\end{array}\right].
\end{align}
The logit transformation is used here because both $\beta$ and $\gamma$ are rates which are bounded from zero to one, but the GP prior has positive measures over the negative reals. Other transformation, such as the probit function, $\Phi^{-1}(x)$, the cumulative log-log function, $\log(-\log(x))$, or the identity function, $x$, could be also used here. $\mu_j(\cdot)$ is the mean function, where we assume a constant mean, i.e., $\mu_j(\mathbf{x})=\mu_j$. $\tau>0$ is the process variance, and $K_{\boldsymbol{\phi}_j}$ is the correlation function, for which a Gaussian correlation function is commonly used in the form of $K_{\boldsymbol{\phi}_j}(\mathbf{x},\mathbf{x}')=\exp(-\|\boldsymbol{\phi}_j\odot(\mathbf{x}-\mathbf{x}')\|_2^2)$ for any $\mathbf{x},\mathbf{x}'\in\Omega\subseteq\mathbb{R}^d$, where $\boldsymbol{\phi}_j\in\mathbb{R}^d$ is the unknown lengthscale parameter and $\odot$ denotes the element-wise product of two vectors. Note that the correlation function is usually reparameterized as 
\begin{equation}\label{eq:kernel}
    K_{\boldsymbol{\phi}_j}(\mathbf{x},\mathbf{x}')=\prod^d_{l=1}\phi_{jl}^{4(x_{l}-x'_{l})^2} \quad\text{for any}\quad\mathbf{x},\mathbf{x}'\in\Omega,
\end{equation}
where $\boldsymbol{\phi}_j=(\phi_{j1},\ldots,\phi_{jd})\in(0,1)^d$, for the purpose of  numerical stability, because the domain of $\phi_{jl}\in(0,1)$ is now bounded. See, for example, \cite{brown2018nonparametric} and \cite{mak2017efficient}. As a result, the form of the kernel function \eqref{eq:kernel} is used throughout this article.

In \eqref{eq:GPassumption}, we assume that $\text{logit}(\beta(\cdot))$ and $\text{logit}(\gamma(\cdot))$ are correlated with a positive \textit{cross-correlation}, $0<\rho\leq 1$, which implies that for any given $\mathbf{x}\in\Omega$, the correlation between $\text{logit}(\beta(\mathbf{x}))$ and $\text{logit}(\gamma(\mathbf{x}))$ is $\rho$. This can be verified by \eqref{eq:GPassumption_correlation} and the fact that $K_{\boldsymbol{\phi}_1}(\mathbf{x},\mathbf{x})=K_{\boldsymbol{\phi}_2}(\mathbf{x},\mathbf{x})=1$ for any $\mathbf{x}\in\Omega$. The dependence assumption of the two parameters, $\beta(\cdot)$ and $\gamma(\cdot)$,  is crucial and appealing from an epidemiological perspective. For the compartmental models like SIR, it is well known that the parameters are strongly \textit{coupled} in the modeling literature. See, for example, the joint posterior distribution in \cite{roda2020difficult} which shows that the two parameters in an SIR model are highly positively correlated. Thus, the independent GP assumption as in \cite{brown2018nonparametric} and \cite{plumlee2016calibrating} is not valid in this application. Note that unlike the covariance structures in \cite{banerjee2002prediction,qian2008gaussian} where the parameters $\boldsymbol{\phi}_1$ and $\boldsymbol{\phi}_2$ are assumed to be identical, the covariance structure \eqref{eq:GPassumption} adopts the one in \cite{fricker2013multivariate} and \cite{svenson2016multiobjective}, which is more flexible as the two lengthscale parameters are not necessarily identical. 

Lastly, the fraction $\kappa(t)$ is assumed to have a Gaussian process prior:
\begin{equation}\label{eq:kappa}
    {\rm logit }\,\kappa(\cdot)\sim\mathcal{GP}(\mu_3,\nu K_{\varphi}(\cdot,\cdot))
\end{equation}
with $\nu>0$ and $\varphi\in(0,1)$, where $K_{\varphi}$ has the same form of \eqref{eq:kernel}.

Suppose that we observe the reported infectious cases in $n$ days, which are denoted by $\mathbf{y}_n=(y_1,\ldots,y_n)$. Denote $\kappa_t=\kappa(t),\beta_t=\beta(\mathbf{x}_t)$, $\gamma_t=\gamma(\mathbf{x}_t)$,  $\boldsymbol{\kappa}=(\kappa_1,\ldots,\kappa_n),\boldsymbol{\beta}=(\beta_1,\ldots,\beta_n)$, and $\boldsymbol{\gamma}=(\gamma_1,\ldots,\gamma_n)$. Furthermore, denote $\mathbf{1}_n=(1,\ldots,1)^T\in\mathbb{R}^{n\times 1},\mathbf{K}_{\boldsymbol{\phi}_j}=(K_{\boldsymbol{\phi}_j}(\mathbf{x}_i,\mathbf{x}_k))_{1\leq i,k\leq n}\in\mathbb{R}^{n\times n}$, $\mathbf{K}_{\varphi}=(K_{\varphi}(i,k))_{1\leq i,k\leq n}\in\mathbb{R}^{n\times n}$, and denote $a_{ij}$ as the element $\mathbf{A}_{ij}$ . Then, together with the model assumptions \eqref{eq:dataassumption}, \eqref{eq:GPassumption}, \eqref{eq:kernel} and \eqref{eq:kappa}, we have the following hierarchical model,
\begin{align}
    y_t|\boldsymbol{\beta},\boldsymbol{\gamma},\boldsymbol{\kappa}&\overset{\text{indep.}}{\sim} \text{Poi}(\kappa_tf(t,\beta_t,\gamma_t))\quad\text{for}\quad t=1,\ldots,n,\nonumber\\
{\rm logit }\,\left(
\begin{array}{c}
\boldsymbol{\beta}\\
\boldsymbol{\gamma}\\
\end{array}\right)&\sim\mathcal{N}_{2n}\left(\left[
\begin{array}{c}
\mu_1\mathbf{1}_n\\
\mu_2\mathbf{1}_n\\
\end{array}\right], \tau\boldsymbol{\Sigma}\right),\nonumber\\
{\rm logit }\,(\boldsymbol{\kappa})&\sim\mathcal{N}_n(\mu_3\mathbf{1}_n,\nu\mathbf{K}_{\varphi}),\nonumber\\
\tau&\sim\text{InvGamma}(a_{\tau},b_{\tau}),\label{eq:prior1}\\
\rho&\sim\text{Beta}(1,b_{\rho}),\label{eq:prior2}\\
\nu&\sim\text{InvGamma}(a_{\nu},b_{\nu}),\label{eq:prior3}\\
\varphi&\sim\text{Beta}(1,b_{\varphi}),\label{eq:prior4}\\
\phi_{jl}&\overset{\text{indep.}}{\sim}\text{Beta}(1,b_{\phi})\quad\text{for}\quad j=1,2;l=1,\ldots,d,\label{eq:prior5}\\
\mu_j&\overset{\text{indep.}}{\sim}\mathcal{N}(\alpha_j,\sigma^2_j)\label{eq:prior6}\quad\text{for}\quad j=1,2,3,
\end{align}
where 
$$
\boldsymbol{\Sigma}=\left[
\begin{array}{cc}
a^2_{11}\mathbf{K}_{\boldsymbol{\phi}_1} + a^2_{12}\mathbf{K}_{\boldsymbol{\phi}_2}&a_{11}a_{21}\mathbf{K}_{\boldsymbol{\phi}_1} + a_{12}a_{22}\mathbf{K}_{\boldsymbol{\phi}_2}\\
a_{11}a_{21}\mathbf{K}_{\boldsymbol{\phi}_1} + a_{12}a_{22}\mathbf{K}_{\boldsymbol{\phi}_2}&a^2_{21}\mathbf{K}_{\boldsymbol{\phi}_1} + a^2_{22}\mathbf{K}_{\boldsymbol{\phi}_2}\\
\end{array}\right],
$$
and \eqref{eq:prior1}, \eqref{eq:prior2}, \eqref{eq:prior3}, \eqref{eq:prior4}, \eqref{eq:prior5}, \eqref{eq:prior6} are the priors of the parameters, in which  InvGamma$(a,b)$ is an inverse gamma distribution with shape parameter $a$ and rate parameter $b$, and  Beta$(1,b)$ is a beta distribution with parameters 1 and $b$. 

\subsection{Posterior Distributions}\label{sec:posterior}
The goal of this study is to infer the functional parameters $\beta(\mathbf{x})$ and $\gamma(\mathbf{x})$ and subsequently investigate whether the $d$-dimensional factor $\mathbf{x}$ plays a role in varying the basic reproduction number, which is denoted by $\mathcal{R}_0(\mathbf{x}):=\beta(\mathbf{x})/\gamma(\mathbf{x})$. In addition, predicting the number of future infections based on forecast weather and government interventions, say $\mathbf{x}_{n+1}$, is also of great interest. Therefore, the joint posterior distribution of $\beta(\mathbf{x}),\gamma(\mathbf{x})$, and the number of the future daily infected cases are developed as follows. 

We first derive the joint posterior distribution of $\beta(\mathbf{x})$ and $\gamma(\mathbf{x})$. Denote the parameters  $\boldsymbol{\psi}=(\tau,\rho,\nu,\varphi, \boldsymbol{\phi}_1,\boldsymbol{\phi}_2,\mu_1,\mu_2)$ and $\text{data}=\{y_t,\mathbf{x}_t\}_{1\leq t\leq n}$. Then, the posterior distribution given observations can be obtained by
\begin{align*}
    \pi(\beta(\mathbf{x}),\gamma(\mathbf{x}),&\boldsymbol{\beta},\boldsymbol{\gamma},\boldsymbol{\kappa},\boldsymbol{\psi}|\text{data})\\
    \propto&\pi(\beta(\mathbf{x}),\gamma(\mathbf{x})|\boldsymbol{\beta},\boldsymbol{\gamma},\boldsymbol{\kappa},\boldsymbol{\psi},\text{data})\pi(\boldsymbol{\beta},\boldsymbol{\gamma},\boldsymbol{\kappa},\boldsymbol{\psi}|\text{data}),
\end{align*}
where $\pi(x|y)$ denotes the posterior distribution of $x$ given $y$. Thus, the joint posterior distribution of $\beta(\mathbf{x})$ and $\gamma(\mathbf{x})$ can be approximated by Markov chain Monte Carlo (MCMC) by drawing the samples from $\pi(\beta(\mathbf{x}),\gamma(\mathbf{x})|\boldsymbol{\beta},\boldsymbol{\gamma},\boldsymbol{\kappa},\boldsymbol{\psi},\text{data})$ and $\pi(\boldsymbol{\beta},\boldsymbol{\gamma},\boldsymbol{\kappa},\boldsymbol{\psi}|\text{data})$ iteratively. The posterior $\pi(\beta(\mathbf{x}),\gamma(\mathbf{x})|\boldsymbol{\beta},\boldsymbol{\gamma},\boldsymbol{\kappa},\boldsymbol{\psi},\text{data})$ can be drawn based on the property of conditional multivariate normal distributions, that is,
\begin{align}\label{eq:multinormal}
    &{\rm logit }\,\left(
\begin{array}{c}
\beta(\mathbf{x})\\
\gamma(\mathbf{x})\\
\end{array}\right)|\boldsymbol{\beta},\boldsymbol{\gamma},\boldsymbol{\kappa},\boldsymbol{\psi},{\rm{data}}
\sim \mathcal{N}_2 \left(u(\mathbf{x}),s(\mathbf{x}) \right),
\end{align}
where
$$
u(\mathbf{x})=\left[\begin{array}{c}
         \mu_1  \\
       \mu_2
    \end{array}\right]+\Sigma(\mathbf{x})\boldsymbol{\Sigma}^{-1}\left(
\begin{array}{c}
{\rm logit }(\boldsymbol{\beta})-\mu_1\mathbf{1}_n\\
{\rm logit }(\boldsymbol{\gamma})-\mu_2\mathbf{1}_n\\
\end{array}\right)
$$
and 
$$
s(\mathbf{x})=\tau\left(\left[
\begin{array}{cc}
1&\rho\\
\rho&1\\
\end{array}\right]-\Sigma(\mathbf{x})\boldsymbol{\Sigma}^{-1}\Sigma(\mathbf{x})^T\right)
$$
with 
$$
\Sigma(\mathbf{x})=\left[
\begin{array}{cc}
a^2_{11}\mathbf{k}_{\boldsymbol{\phi}_1}(\mathbf{x}) + a^2_{12}\mathbf{k}_{\boldsymbol{\phi}_2}(\mathbf{x})&a_{11}a_{21}\mathbf{k}_{\boldsymbol{\phi}_1}(\mathbf{x}) + a_{12}a_{22}\mathbf{k}_{\boldsymbol{\phi}_2}(\mathbf{x})\\
a_{11}a_{21}\mathbf{k}_{\boldsymbol{\phi}_1}(\mathbf{x}) + a_{12}a_{22}\mathbf{k}_{\boldsymbol{\phi}_2}(\mathbf{x})&a^2_{21}\mathbf{k}_{\boldsymbol{\phi}_1}(\mathbf{x}) + a^2_{22}\mathbf{k}_{\boldsymbol{\phi}_2}(\mathbf{x})\\
\end{array}\right],
$$
where $\mathbf{k}_{\boldsymbol{\phi}_j}(\mathbf{x})=(K_{\boldsymbol{\phi}_j}(\mathbf{x},\mathbf{x}_1),\ldots,K_{\boldsymbol{\phi}_j}(\mathbf{x},\mathbf{x}_n))$. The MCMC samples of $\beta(\mathbf{x})$ and $\gamma(\mathbf{x})$ can then be obtained by sampling from the multivariate normal distribution of \eqref{eq:multinormal} and taking the inverse of the logit function. For the posterior $\pi(\boldsymbol{\beta},\boldsymbol{\gamma},\boldsymbol{\kappa},\boldsymbol{\psi}|\text{data})$, we have 
\begin{align}\label{eq:jointdis}
 &\pi(\boldsymbol{\beta},\boldsymbol{\gamma},\boldsymbol{\kappa},\boldsymbol{\psi}|\rm{data})\propto\pi({\rm{data}}|\boldsymbol{\beta},\boldsymbol{\gamma},\boldsymbol{\kappa},\boldsymbol{\psi})\pi(\boldsymbol{\beta},\boldsymbol{\gamma}|\boldsymbol{\kappa},\boldsymbol{\psi})\pi(\boldsymbol{\kappa}|\boldsymbol{\psi})\pi(\boldsymbol{\psi})\\
 &\propto\exp\left\{-\sum^n_{t=1}\kappa_tf(t,\beta_t,\gamma_t)\right\}\times\prod^n_{t=1}\kappa^{y_t}_tf(t,\beta_t,\gamma_t)^{y_t}\nonumber\\
 &\times\exp\left\{-\frac{1}{\tau}\left(
\begin{array}{c}
{\rm logit }(\boldsymbol{\beta})-\mu_1\mathbf{1}_n\\
{\rm logit }(\boldsymbol{\gamma})-\mu_2\mathbf{1}_n\\
\end{array}\right)^T\boldsymbol{\Sigma}^{-1}\left(
\begin{array}{c}
{\rm logit }(\boldsymbol{\beta})-\mu_1\mathbf{1}_n\\
{\rm logit }(\boldsymbol{\gamma})-\mu_2\mathbf{1}_n\\
\end{array}\right)\right\}\nonumber\\
&\times\exp\left\{-\frac{1}{\nu}\left(
{\rm logit }(\boldsymbol{\kappa})-\mu_3\mathbf{1}_n\right)^T\mathbf{K}_{\varphi}^{-1}\left(
{\rm logit }(\boldsymbol{\kappa})-\mu_3\mathbf{1}_n\right)\right\}\nonumber\\
&\times|\boldsymbol{\Sigma}|^{-1}|\mathbf{K}_{\varphi}|^{-1} \tau^{n+a_{\tau}-1}\exp\{-b_{\tau}\tau\}(1-\rho)^{b_{\rho}-1}\nu^{n/2+a_{\nu}-1}\exp\{-b_{\nu}\nu\}\nonumber\\
&\times\varphi^{b_{\varphi}-1}\prod^2_{j=1}\prod^d_{l=1}(1-\phi_{jl})^{b_{\phi}-1}\exp\left\{-\frac{1}{2}\sum^2_{j=1}\frac{\left(\mu_j-\alpha_j\right)^2}{\sigma_j^2} \right\}.\nonumber
\end{align}
The samples from this posterior distribution can be drawn by Gibbs sampling with Metropolis-Hastings algorithm, the details of which are given in Section 1 of the Supplementary Material \citep{suppsung2022}.

Now we move to the posterior distribution of the number of future infections. Let $\mathbf{x}_{n+1}$ be the forecast weather and government interventions at time $n+1$, and denote $\beta_{n+1}=\beta(\mathbf{x}_{n+1}),\gamma_{n+1}=\gamma(\mathbf{x}_{n+1}),\kappa_{n+1}=\kappa(n+1)$. Then the posterior distribution of the infected number, $y_{n+1}$, given the current observed data has
\begin{align*}
    \pi(y_{n+1},&\beta_{n+1},\gamma_{n+1},\kappa_{n+1},\boldsymbol{\beta},\boldsymbol{\gamma},\boldsymbol{\kappa},\boldsymbol{\psi}|\mathbf{x}_{n+1},\text{data})\\
    \propto&\pi(y_{n+1}|\beta_{n+1},\gamma_{n+1},\kappa_{n+1})\pi(\beta_{n+1},\gamma_{n+1}|\boldsymbol{\beta},\boldsymbol{\gamma},\boldsymbol{\kappa},\boldsymbol{\psi},\mathbf{x}_{n+1})\\
   \times &\pi(\kappa_{n+1}|\boldsymbol{\beta},\boldsymbol{\gamma},\boldsymbol{\kappa},\boldsymbol{\psi})\pi(\boldsymbol{\beta},\boldsymbol{\gamma},\boldsymbol{\kappa},\boldsymbol{\psi}|\text{data}).
\end{align*}
Similarly, this posterior distribution can be drawn by MCMC sampling, where the samples for $\pi(\boldsymbol{\beta},\boldsymbol{\gamma},\boldsymbol{\kappa},\boldsymbol{\psi}|\text{data})$ can be drawn as introduced before, and the samples from $\pi(\beta_{n+1},\gamma_{n+1}|\boldsymbol{\beta},\boldsymbol{\gamma},\boldsymbol{\kappa},\boldsymbol{\psi},\mathbf{x}_{n+1})$ can be similarly drawn from the multivariate normal distribution \eqref{eq:multinormal}. The samples of $\kappa_{n+1}$ from $\pi(\kappa_{n+1}|\boldsymbol{\beta},\boldsymbol{\gamma},\boldsymbol{\kappa},\boldsymbol{\psi})$ can be drawn from the posterior of $\kappa(t)$,
\begin{align}\label{eq:kappat}
    {\rm logit }\,(\kappa(t))|\boldsymbol{\beta},\boldsymbol{\gamma},\boldsymbol{\kappa},\boldsymbol{\psi},
\sim \mathcal{N} (\mu_3+\mathbf{k}_{\varphi}(t)\mathbf{K}_{\varphi}^{-1}({\rm logit }\,(\boldsymbol{\kappa})-\mu_3\mathbf{1}_n),\\
 \nu(1-\mathbf{k}_{\varphi}(t)\mathbf{K}_{\varphi}^{-1}\mathbf{k}_{\varphi}(t)^T)),\nonumber
\end{align}
and set $t=n+1$, where $\mathbf{k}_{\varphi}(t)=(K_{\varphi}(t,1),\ldots,K_{\varphi}(t,n))$. The distribution, $\pi(y_{n+1}|\beta_{n+1},\gamma_{n+1},\kappa_{n+1})$, follows a Poisson distribution with the mean $\kappa_{n+1}f(n+1,\beta_{n+1},\gamma_{n+1})$. Thus, the MCMC samples can be drawn iteratively from these posteriors.

\section{Simulation Study}\label{sec:simulation}
In this section, simulation studies are conducted to examine the performance of the proposed method.  In the simulations, the hyperparameters in the priors \eqref{eq:prior1}, \eqref{eq:prior2}, \eqref{eq:prior3}, \eqref{eq:prior4}, \eqref{eq:prior5}, \eqref{eq:prior6}, are set as follow. Similar to \cite{brown2018nonparametric}, the shape parameters $b_{\rho},b_{\phi}$ and $b_{\varphi}$ are chosen to be 0.1, which place most probability mass near one to enforce the smoothness for the functional parameters; $a_{\tau}=a_{\nu}=0.01$ and $b_{\tau}=b_{\nu}=0.01$ are chosen so that the prior is centered at one with standard deviation $\sqrt{0.01/0.01^2}=10$; for \eqref{eq:prior6} we set $\alpha_j=0$ and $\sigma^2_j=1$ for $j=1,2,3$. For the MCMC sampling, 2,000 iterations are performed in a burn-in period, and after that additional 2,000 MCMC samples are drawn, which are thinned to reduce autocorrelation.

\begin{figure}[h!]
    \centering
    \includegraphics[width=\textwidth]{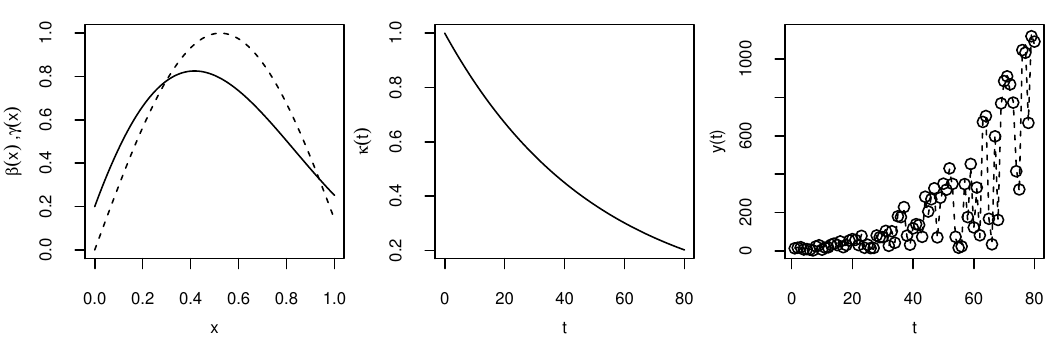}
    \caption{Simulation setting. The left panel demonstrates $\beta(x)$ (solid line) and $\gamma(x)$ (dashed line), the middle panel demonstrates $\kappa(t)$, and the right panel demonstrates the true mean function,  $\kappa(t)f(t,\beta(x), \gamma(x))$ (dashed line), and the simulated data (dots).}
    \label{fig:simulation_setup}
\end{figure}

Suppose that the observation $y_t$ is simulated from a Poisson distribution with the mean function, $\kappa(t)f(t,\beta(x), \gamma(x))$, where $f(t,\beta(x), \gamma(x))=(t/10+5\beta(x)+\gamma(x)(t/10)^2)^2$ and $x$ is one-dimensional factor in the space $[0,1]$. Let $\beta(x)=\sin(3x)\exp(-x)+0.2,\gamma(x)=\sin(3x)$, and $\kappa(t)=\exp(-t/50)$, which are demonstrated in the left and middle panels of Figure \ref{fig:simulation_setup}. It can be seen that the two curves, $\beta(x)$ and $\gamma(x)$, share some similarity overall, which suggests that the dependence assumption of these two functions is necessary. We generate $x_1,\ldots,x_{80}$ from a uniform distribution, and randomly generate $n=80$ observations, $y_1,\ldots,y_{80}$. The right panel of Figure \ref{fig:simulation_setup} shows the random samples as dots, where the solid line is the true mean function, $\kappa(t)f(t,\beta(x), \gamma(x))$. We use the first 60 samples, $y_1,\ldots,y_{60}$, as the training dataset and the other 20 samples as the test dataset. 

\begin{figure}[h!]
    \centering
    \includegraphics[width=\textwidth]{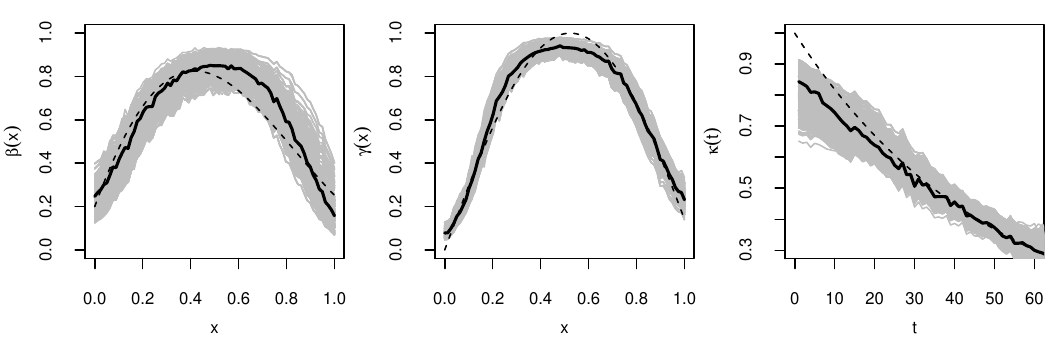}
    \caption{Posteriors of $\beta(x)$ (left panel), $\gamma(x)$ (middle panel), and $\kappa(t)$ (right panel). The gray lines are the MCMC draws, the solid lines are the posterior mean, and the dashed lines are the true functions. }
    \label{fig:simulation_correlation}
\end{figure}

\begin{figure}[h!]
    \centering
    \includegraphics[width=0.6\textwidth]{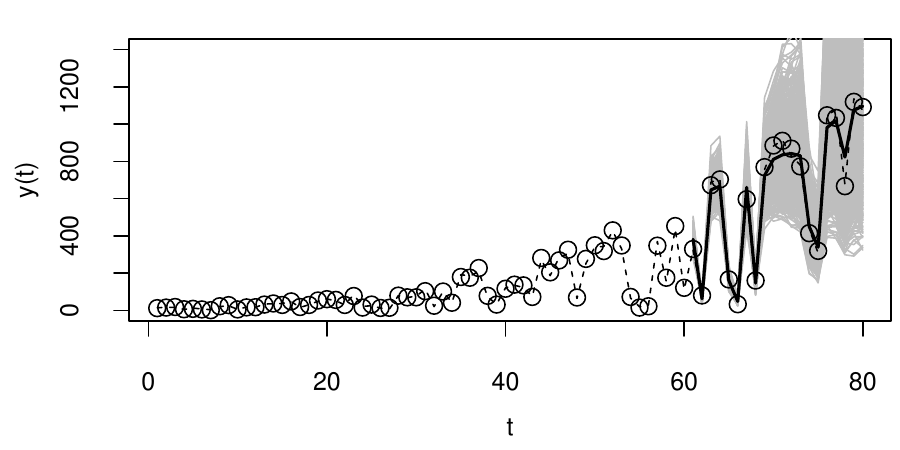}
    \caption{Posteriors of test data $y_{61},\ldots,y_{80}$. The gray lines are the MCMC draws, the solid lines are the posterior mean, and the dashed lines are the true functions.     }
    \label{fig:simulation_prediction}
\end{figure}

Figure \ref{fig:simulation_correlation} shows the posterior draws of $\beta(x),\gamma(x)$ and $\kappa(t)$. It can be seen that the posterior means can recover the true functions very well. The predictions on the test dataset are presented in  Figure \ref{fig:simulation_prediction}, which shows that the posterior mean is reasonably close to the true function. These results demonstrate that the proposed method can perform well for the models with functional parameters in terms of estimation and prediction.

\section{Application to COVID-19}\label{sec:application}
In this section, we use the proposed model to analyze the COVID-19 virus spread among the eight largest metropolitan areas in the United States (US). In particular, we use the model to estimate the impacts of weather and government interventions (and their interactions) on virus transmissibility, to forecast daily infected cases based on these factors, and to estimate the fraction of cases reported.

\begin{figure}[t!]
    \centering
    \includegraphics[width=0.9\textwidth]{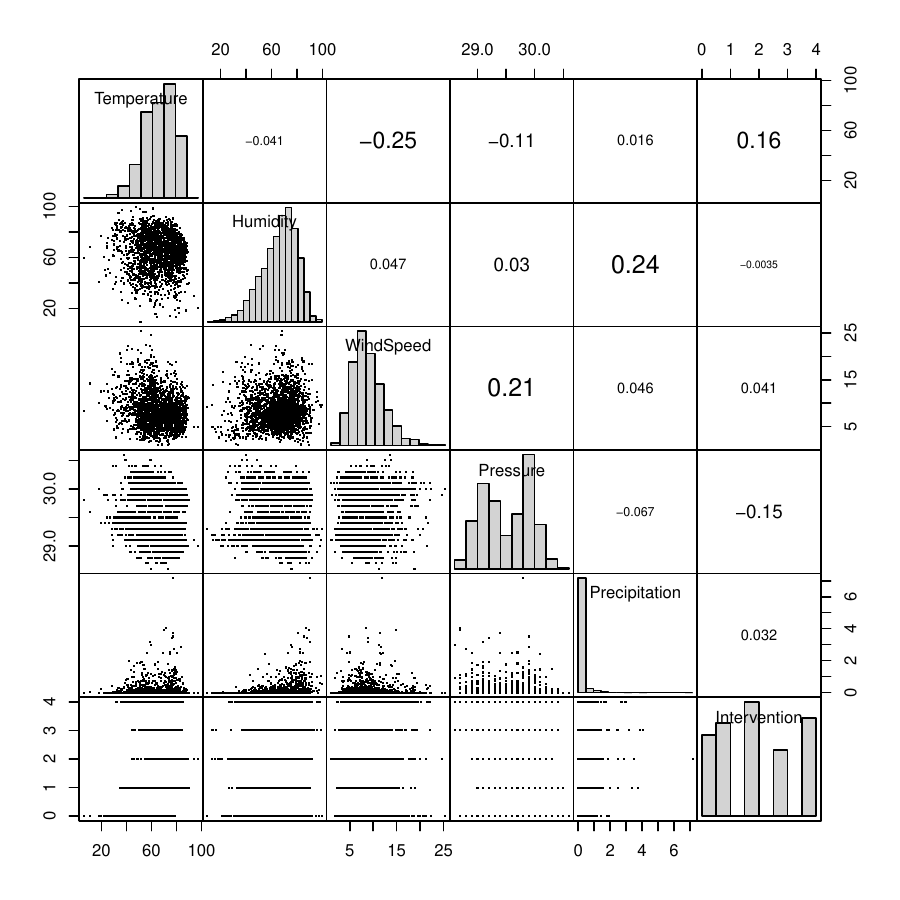}
    \caption{Scatterplots of input factors, where the left diagonal panels are the scatterplots for every pair of factors, the right diagonal panels are the correlations, and the diagnoal panels are the histograms.}
    \label{fig:pairsplot}
\end{figure}

The data source is briefly introduced here. The daily COVID-19 cases are obtained at the US county level from the data repository provided by New York Times \citep{smith2020coronavirus}, starting from January 25, 2020 to  November 25, 2020. The population sizes are obtained from the census bureau website, which also can be found in \cite{yu2020impact}. The historical weather data and the weather forecast are collected from the Weather Underground \citep{wg2020}, which include the daily average temperature, humidity, wind speed, pressure, and precipitation. The information of  government interventions is obtained from New York Times \citep{jasmine2020coronavirus} and local media, where we categorize the interventions into five levels: (0) no intervention; (1) all businesses are open with mask required and some capacity limitations ; (2) all industries resume operations but some indoor services, such as bars and restaurants, remain closed; (3) Industries resume operations with severe restrictions and capacity limitations; (4) all non-essential businesses are closed. Scatterplots for every pair of factors are demonstrated in Figure \ref{fig:pairsplot}, which appears to have no obvious relationship between any pair of the factors.

Now we are ready to apply the proposed model to the data, where the setting of the MCMC sampling is similar to the one in Section \ref{sec:simulation}. Consider the confirmed cases before November 11 as the training data, and the cases from November 11 to 25 as the test data. Since the actual infectious period for COVID-19 is not available and it varies by individual and situation, as suggested by Centers for Disease Control and Prevention and \cite{wilson2020weather},  we assume an infectious period of 11 days from the actual infection to the confirmation of the positive test result.  In other words, we assume that the actual infection occurs 11 days prior to the confirmation date. The input factor is a 6-dimensional variable, i.e., $\mathbf{x}\in \mathbb{R}^6$, including 5 variables representing weather data and one variable representing government intervention levels.  The MCMC samples of the basic reproduction number can be obtained from the MCMC samples of $\beta(\mathbf{x})$ and $\gamma(\mathbf{x})$ by computing $\mathcal{R}_0(\mathbf{x})=\beta(\mathbf{x})/\gamma(\mathbf{x})$. Since it's hard to visualize the function $\mathcal{R}_0(\mathbf{x})$ via a six-dimensional $\mathbf{x}$, similar to \cite{welch1992screening}, we use a functional ANOVA decomposition \citep{Hoeffding1948ACO,sobol1993sensitivity,santner2003design} for $\mathcal{R}_0(\mathbf{x})$ and plot its overall mean and main effects, which respectively measure the overall influence and the influence of a single factor on the basic reproduction number. That is, suppose that $\mathbf{x}$ follows a distribution $F(\mathbf{x})$ where $F(\mathbf{x})=F_1(x_1)\times F_2(x_2)\times\cdots\times F_d(x_d)$, then  the overall mean and the main effects of the function $\mathcal{R}_0(\mathbf{x})$ can be obtained by 
\begin{equation}\label{eq:fanova}
    m_0:=\int_{\Omega}\mathcal{R}_0(\mathbf{x}){\rm{d}}F(\mathbf{x})\quad\text{and}\quad m_j(x_j)=\int_{\Omega_{-j}}(\mathcal{R}_0(\mathbf{x})-m_0){\rm{d}}F_{-j}(\mathbf{x}_{-j}),
\end{equation}
respectively, where $\int_{\Omega_{-j}}\cdots{\rm{d}}F_{-j}(\mathbf{x}_{-j})$ indicates
integration over the variables not in $j$ and $F_{-j}(\mathbf{x}_{-j})=\prod^d_{i\neq j}F_i(x_i)$.  Since the MCMC samples of $\mathcal{R}_0(\mathbf{x})$ are available for any $\mathbf{x}\in\Omega$ and the integration in \eqref{eq:fanova} can be approximated by the Monte-Carlo integration \citep{caflisch1998monte}, the samples of the posterior distributions of $m_0$ and $m_j(x_j)$ can be naturally drawn via a Monte-Carlo sampling method.
This is similar to \cite{le2014bayesian} for estimating the Sobol indices through a surrogate model that accounts for both the integration errors and the surrogate model uncertainty.

\begin{figure}[h!]
    \centering
    \includegraphics[width=0.6\textwidth]{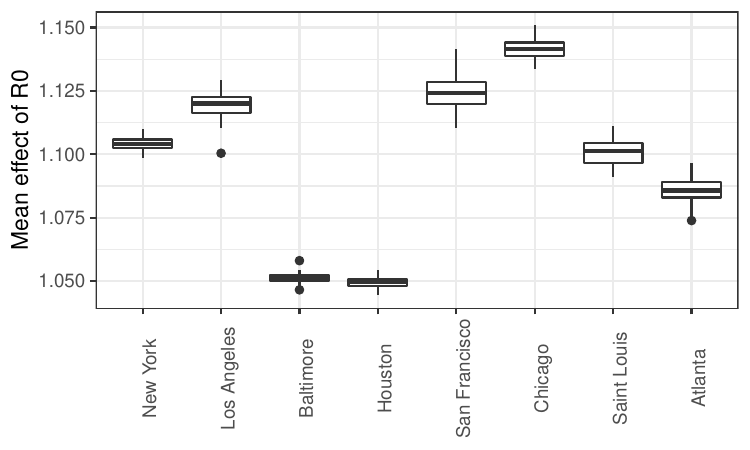}
    \caption{Overall mean of basic reproduction number.}
    \label{fig:mean_effect}
\end{figure}

\begin{figure}[h!]
    \centering
    \includegraphics[width=\textwidth]{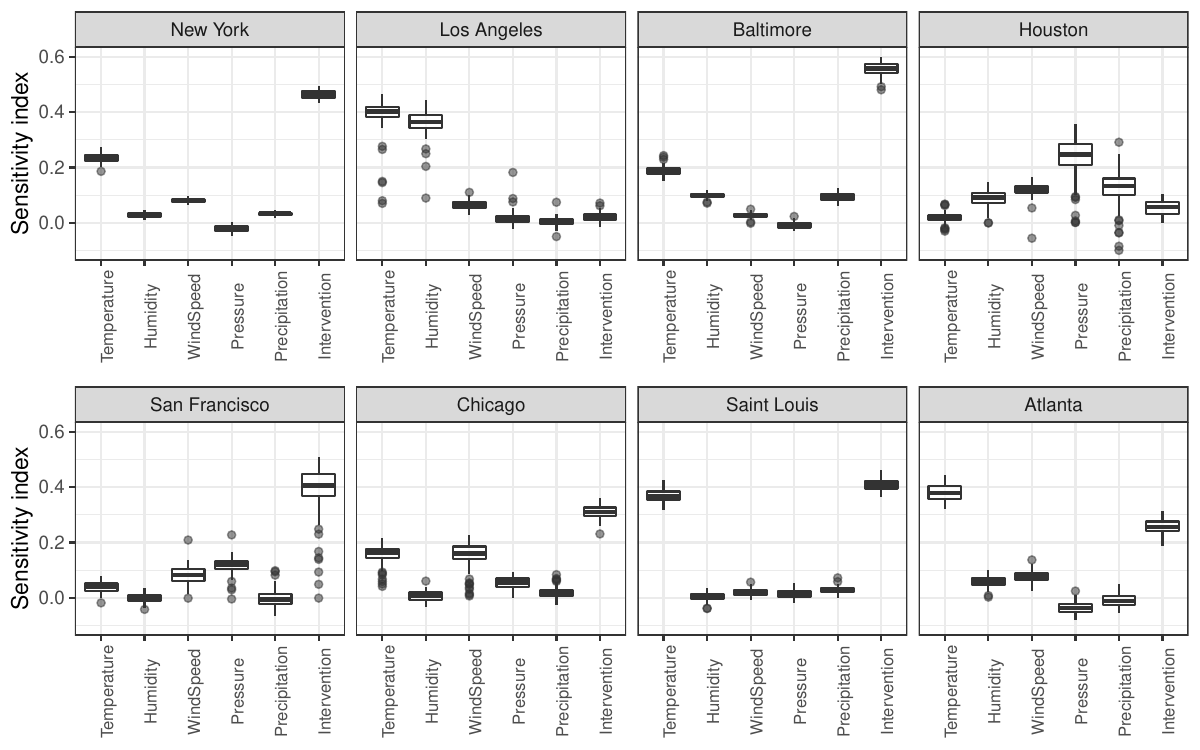}
    \caption{Main effect indices of basic reproduction number.}
    \label{fig:main_effect_sa}
\end{figure}

The boxplots of the overall means of $\mathcal{R}_0(\mathbf{x})$ are shown in Figure \ref{fig:mean_effect}. It can be seen that among these eight cities, Chicago has the highest basic reproduction number than other cities, which implies that each existing infection in Chicago can cause more new infections than other cities, while the existing infection in Baltimore and Houston causes fewer new infections. Before illustrating the main effects, sensitivity analysis \citep{sobol1993sensitivity} is adopted to  determine which input factors are responsible for the most variation in the basic reproduction number. The result is shown in Figure \ref{fig:main_effect_sa}. Although no unique factor can dominate the others for all of the cities in terms of sensitivity index, it appears that government interventions have made stronger impacts than weather factors on the virus spread in most of the cities, especially in  Baltimore and San Francisco. On the other hand, some cities, such as Los Angeles, Saint Louis, and Atlanta, have shown evidence that temperature has played a crucial role in explaining the variation of the basic reproduction number.

\begin{figure}[h!]
    \centering
    \includegraphics[width=\textwidth]{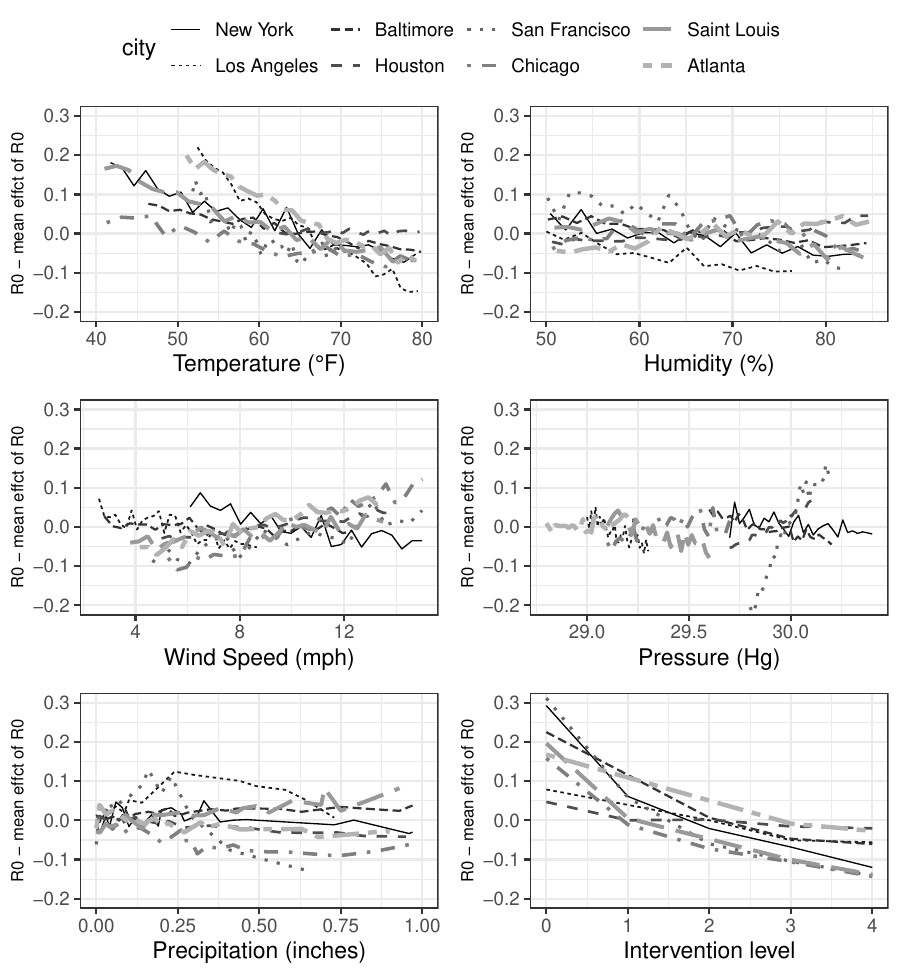}
    \caption{Main effect of basic reproduction number, $m_j(x_j)$, which illustrates the influence of a single factor on the reproduction number.}
    \label{fig:main_effect}
\end{figure}

The main effects of $\mathcal{R}_0(\mathbf{x})$ are demonstrated in Figure \ref{fig:main_effect}. As shown in the sensitivity analysis, the intervention and temperature factors both have larger variations in the main effects, ranging from -0.1 to 0.3, whereas the main effects of other weather factors mostly range from -0.1 to 0.1. Among these six factors, it shows that  temperature and government intervention both have negative effects on the virus spread for all of the cities, while other factors have no common trend. In particular, it appears that a decrease of 10$^\circ$F in temperature leads to an increase of roughly 0.06 (with a standard error (SE) of 0.0354) in the basic reproduction number. This result is quite promising in the sense that  most of the existing methods cannot directly quantify the effect of temperature on the basic reproduction number. The intervention factor shows that the basic reproduction number can be effectively reduced if governments implement more restrictions to combat the COVID-19 outbreak, especially for New York and San Francisco, where a change from no intervention to the strictest restrictions can lead to a decrease in the basic reproduction number of approximately 0.42 (SE 0.03). This finding is consistent with the results in some recent work on the effect of government intervention for COVID-19, such as \cite{flaxman2020estimating,haug2020ranking,haldar2020effect,wang2020mitigate}.

\begin{figure}[t]
    \centering
    \includegraphics[width=\textwidth]{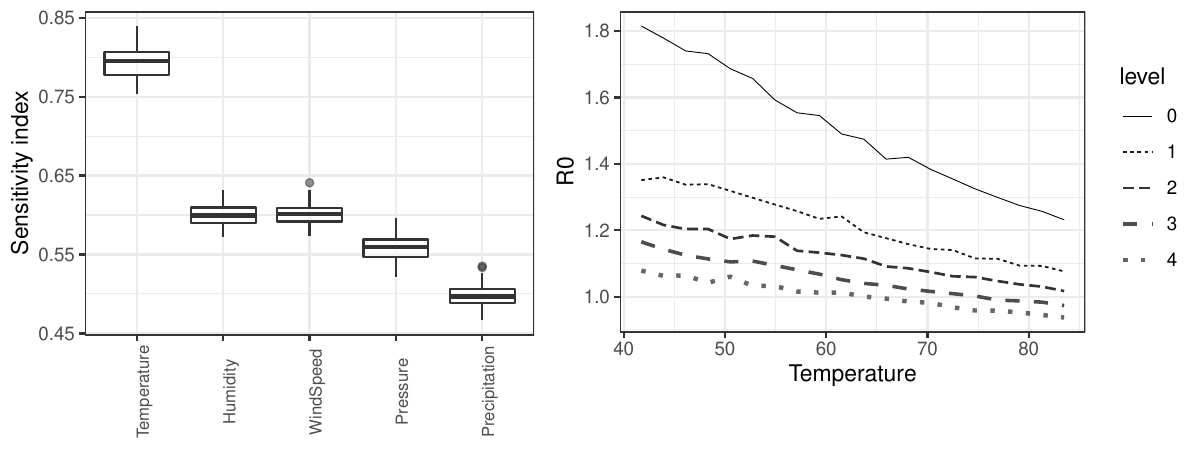}
    \caption{Interaction effect indices (left) and interaction plot (right) between temperature and government intervention for New York.}
    \label{fig:interaction_effect_plot}
\end{figure}

We further investigate the interaction effects of the basic reproduction number. Particularly, we focus on the interaction effects between the  intervention factor, which is controllable by governments, and the five weather factors, which are uncontrollable. The sensitivity indices of these five interaction effects are first computed to compare their relative importance. For the sake of saving the space, only the interaction effects for New York are demonstrated here. The sensitivity indices and the interaction plot with the highest index are respectively shown in the left and right panels of Figure \ref{fig:interaction_effect_plot}. It can be seen that the interaction effect between temperature and government interventions has the highest sensitivity index, and from the interaction plot of the two factors, it appears that when governments implement more restrictions, the effect of temperature on the virus spread tends to be milder. This result suggests that as the weather gets colder, policymakers may need to implement more restrictions to mitigate the pandemic.

\begin{figure}[h!]
    \centering
    \includegraphics[width=\textwidth]{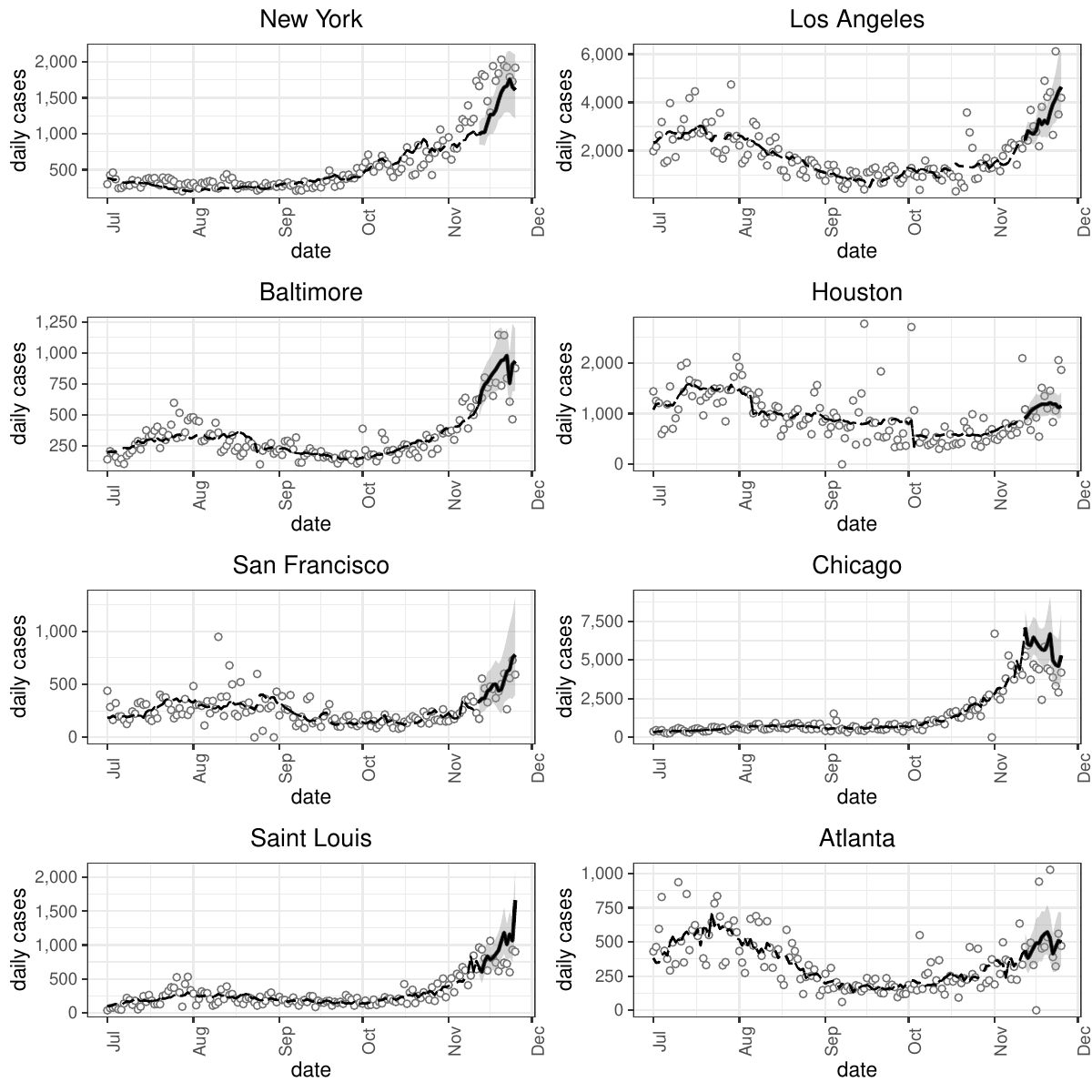}
    \caption{Prediction performance of the proposed model. The dots are the confirmed numbers, the dashed lines are the fitted values from July 1 to November 11, the gray lines and the solid lines are the MCMC draws and posterior means for the test data from November 12 to 25.}
    \label{fig:newcase_prediction}
\end{figure}

We validate the proposed model by performing predictions on the test data from November 12 to 25. The prediction results of the eight cities are shown in Figure \ref{fig:newcase_prediction}. The predictions are reasonably accurate over the 14-day period. Particularly, in the cities New York, Los Angeles, Baltimore, and San Francisco, the infected cases tend to increase over the 14-day period and our predictions successfully capture the trend. This shows strong empirical justification for our model specification.

\begin{figure}[h!]
    \centering
    \includegraphics[width=\textwidth]{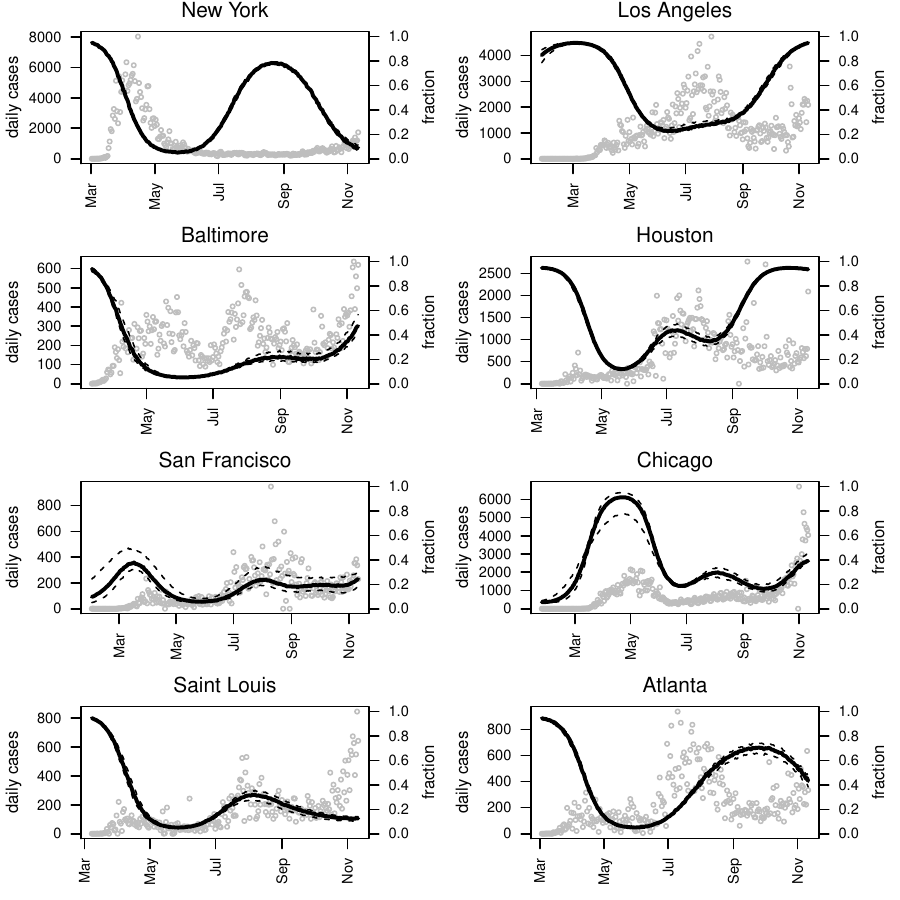}
    \caption{The posteriors of the fraction of cases reported, $\kappa(t)$, where the solid and dashed lines are the median estimate and 95\% confidence intervals, respectively, and the gray dots are the daily confirmed cases.}
    \label{fig:realdata_fraction}
\end{figure}

Figure \ref{fig:realdata_fraction} presents the posteriors of the fraction of cases reported, $\kappa(t)$. The posteriors show that the actual infections are greatly undetected in most of the cities.  This finding coincides with the recent results by \cite{CDConline,pei2021burden,noh2021estimation}. In particular, San Francisco shows that the fraction less than 40\% for the entire time of the ongoing pandemic, and New York shows low fractions during the peak of confirmed cases, which suggests that the number of actual infections during the peak is  likely much higher (over 20,000 daily cases) than the confirmed daily cases (about 8,000). The estimation of the fractions provides crucial insights for public health, which  determines the actual severity of COVID-19 and can be used to develop effective strategies against the outbreak.

\begin{figure}[h!]
    \centering
    \includegraphics[width=0.7\textwidth]{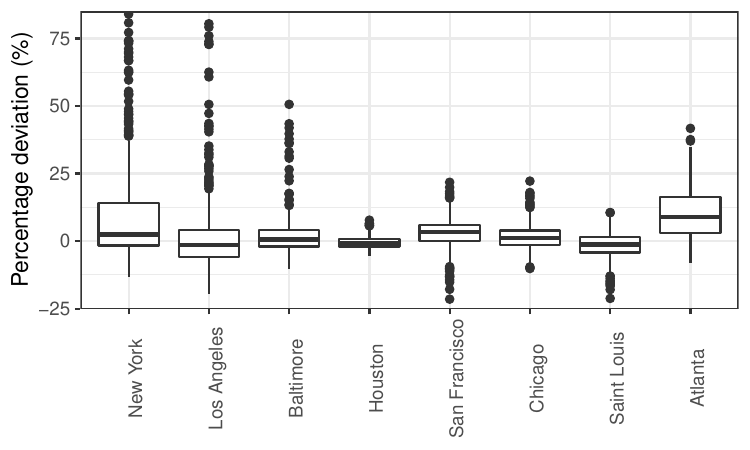}
    \caption{The sensitivity analysis for the priors across 30 different prior specifications, which shows the percentage deviations between the original posteriors and the posteriors with the alternative priors.}
    \label{fig:deviation}
\end{figure}

To examine the robustness of the results, sensitivity analysis for the priors is conducted, which  examines  the impact of 30 different prior specifications on the resulting posteriors of $\mathcal{R}_0(\mathbf{x})$ for the test data. The percentage deviations in the average posterior estimates between the original model and the models with the 30 alternative prior specifications are computed, which are presented in Figure \ref{fig:deviation}. It appears that the results are somewhat stable under different prior settings, with the average percentage deviation being about 3.6\%. The posteriors based on the data of New York city are found to have higher shifts with some prior settings, indicating that it may require a bit more care in the future analysis. 

\section{Concluding Remarks}\label{sec:conclusion}
How the weather and government interventions affect the spread of a disease has been an important question but remains unclear in the literature. A new statistical model incorporated with the prominent SIR model is employed to study the impact on the COVID-19 virus transmissibility among eight US metropolitan areas. Gaussian process modeling and sensitivity analysis for the functional parameters enable to investigate the the main and interaction effects of the factors, which could lead to a new intervention strategy for policymakers. This study shows that, among six factors, temperature and government interventions  have stronger impacts on the COVID-19  spread in most of the cities. The  temperature  has been found to have negative effects in all of the cities. Other weather factors, such wind speed and pressure, do not show common effects among the eight cities. New York city has shown a strong interaction effect between temperature and interventions, which suggests that more restrictions are necessary to mitigate the outbreak as the weather gets colder.

Although we found some potential associations between weather and virus transmissibility, it is worth emphasizing that these associations may not directly imply the causation of the virus transmissibility, meaning that there might be some lurking/causal variables which are correlated with these factors that make the associations appear stronger. For instance, as recent studies have shown (e.g., \cite{wilson2020weather,soucy2020estimating}), the individual mobility may have the direct impact on the COVID-19 spread, which could be strongly correlated with weather factors. Therefore, incorporating the information of individual mobility and estimating the causal effects of mobility and weather is worthwhile to investigate in the future work. In addition, it is conceivable to consider other potential factors for the virus transmission, such as private sector (i.e., non-governmental) interventions, travel restrictions, and public compliance with government recommendations such as vaccinations, quarantines and face coverings, if the data are available. However, including too many factors may lead to over-parameterization which in turn causes unstable results. A potential solution is to perform sensitivity analysis to screen out non-influential factors that have the least effect, and then remove them from the analysis.  We explore these issues in future work.

\begin{acks}[Acknowledgments]

The author gratefully acknowledges the conscientious efforts of the editor and two anonymous reviewers whose comments greatly strengthen this paper. The author is grateful to Dr. Ying Hung for the initial discussion which inspires the author to work on this problem.
\end{acks}

\begin{funding}
The author was supported by NSF Grant DMS-2113407.
\end{funding}
\begin{supplement}
\stitle{Sampling details for the posterior distributions}
\sdescription{The details of sampling for the posterior distributions in Section \ref{sec:posterior} are described in this file.}
\end{supplement}
\begin{supplement}
\stitle{Data and R code}
\sdescription{A zip file containing the data and R code for reproducing the results in Sections \ref{sec:simulation} and \ref{sec:application}.}
\end{supplement}

\bibliographystyle{imsart-nameyear}
\bibliography{bib}

\end{document}